\documentclass[conference]{IEEEtran}
\IEEEoverridecommandlockouts
\usepackage{cite}
\usepackage{amsmath,amssymb,amsfonts}
\usepackage{algorithmic}
\usepackage{graphicx}
\usepackage{textcomp}
\usepackage{xcolor}
\usepackage{hyphenat}

\usepackage[hyphens]{url}
\usepackage[bookmarks=false]{hyperref} 

\usepackage{booktabs}
\usepackage{multirow}
\usepackage{siunitx}

\usepackage{graphics}
\usepackage{xspace}
\usepackage{tabularx}
\usepackage{caption}
\usepackage{subcaption}
\usepackage{float}

\usepackage{url}
\usepackage[left=0.7in,right=0.7in,top=0.7in,bottom=1in]{geometry}

\def\BibTeX{{\rm B\kern-.05em{\sc i\kern-.025em b}\kern-.08em
    T\kern-.1667em\lower.7ex\hbox{E}\kern-.125emX}}
\begin{document}

\title{An Experimental Study of Low-Latency Video Streaming over 5G\\

}


\author{
    \IEEEauthorblockN{
        Imran Khan\IEEEauthorrefmark{1}, Tuyen X. Tran\IEEEauthorrefmark{2}, Matti Hiltunen\IEEEauthorrefmark{2}, Theodore Karagioules\IEEEauthorrefmark{2}, Dimitrios Koutsonikolas\IEEEauthorrefmark{1}
    }
    \IEEEauthorblockA{\IEEEauthorrefmark{1} Northeastern University}
    \IEEEauthorblockA{\IEEEauthorrefmark{2} AT\&T Labs Research}
}

\maketitle

\begin{abstract}
Low-latency video streaming over 5G has become rapidly popular over the last few years due to its increased usage in hosting virtual events, online education, webinars, and all-hands meetings.  Our work aims to address the absence of studies that reveal the real-world behavior of low-latency video streaming. To that end, we provide an experimental methodology and measurements, collected in a US metropolitan area over a commercial 5G network, that correlates application-level QoE and lower-layer metrics on the devices, such as RSRP, RSRQ, handover records, etc., under both static and mobility scenarios. We find that RAN-side information, which is readily available on every cellular device, has the potential to enhance throughput estimation modules of video streaming clients, ultimately making low-latency streaming more resilient against network perturbations and handover events.
\end{abstract}
\vspace{-0.1in}
\section{Introduction}
\label{sec:intro}
Live video streaming traffic is predicted to account for 29.7\% of all Internet traffic by the end of 2023 due to the growing popularity of live video streaming services including live event streaming, shoppable social live streaming, and e-sports and gaming \cite{videostats}. To enable real-time interaction, these applications require minimal end-to-end latency (also known as glass-to-glass latency). Latency is defined as the time delay between video capture and the actual playback at the client, imposing tight content delivery constraints.

In addition, the dynamic nature of cellular connections, due to varying network conditions and user mobility makes it very challenging to ensure a high level of Quality of Experience (QoE) for streaming users.  Video streaming clients are tasked with adapting the requested video bitrate to the available throughput, with the objective of ensuring high-quality uninterrupted streaming. To address the QoE optimization challenge, most video streaming applications typically require accurate throughput estimation, in order to adapt their streaming bitrate to the current network conditions, thus ensuring uninterrupted high-quality streaming. Throughput estimation is typically based on  application layer signals that are updated every few seconds (weighted-averaged throughput probes) and has proven to be sufficient for Video on Demand (VoD) streaming. However, low-latency streaming bearing an additional delivery objective, namely that of timeliness, not only requires more frequent network-condition information updates but is also known to have challenges with conventional throughput estimation techniques, due to chunked transfer encoding.

Concurrently, the latest 5G mobile network technology is being deployed widely in commercial settings, 
promising faster and more reliable connectivity for smartphones, tablets, and other internet-connected devices. It is expected to deliver data speeds up to 20 times faster than 4G networks, with lower latency and higher capacity. 
This, in principle,  opens the door to a new class of latency-critical applications such as low-latency live streaming, augmented reality, connected autonomous vehicles, etc. However, despite these potential benefits, there is a lack of real-world studies investigating the behavior of low-latency video streaming on 5G networks. This paper aims to fill this gap by presenting a study that evaluates the performance of Low-Latency Dash video streaming over a 5G network, providing insights into the feasibility of such applications in real-world settings.



Our study collects client-side QoE metrics using Google's ExoPlayer open-source platform and evaluates performance on 5G networks during both peak and non-peak hours, comparing it to WiFi. Mobility scenarios are also considered, taking into account dynamic network conditions like handovers and low coverage that may arise during user movement. Based on the results of this study, we provide insights into the selection of player configuration and discuss the potential benefit of developing a bitrate adaptation algorithm that uses lower-layer metrics to enhance the user experience of low-latency live streaming.

\vspace{-0.2in}
\section{Background}
\label{sec:background}

MPEG-DASH (Dynamic Adaptive Streaming over HTTP) \cite{MPEG_DASH} is an industry-wide adopted video streaming standard, that is compatible with all popular video codecs such as H.264, H.265, HVEC, and  end-user devices. The typical architecture of the MPEG-DASH client/server system is as follows.  An origin server transcodes the source video content in multiple representations (resolutions/bitrates) and then proceeds to segment each representation into smaller files, namely segments, that are typically 2-10 s in duration. The way the content has been organized in representations and segments is described in a manifest file called an MPD (media presentation description) that is advertised to the video client at the initiation of the streaming session, along with the latency targets set for the particular video stream. The client then proceeds to request every segment sequentially, according to its adaptive bitrate (ABR) module, i.e., an optimization function that decides the representation for every segment. The segments are downloaded at a temporary queue before decoding and playout, known as the buffer. 

DASH has been proven to be ideal for both Video On Demand (VoD) and live streaming, offering high-quality streaming with latency in the range of 10-45 s \cite{wowza}. Nonetheless, to achieve latency in the order of less than 5 s, namely low-latency live streaming, the development of chunked transfer encoding (CTE) with MPEG Common Media Application Format (CMAF) \cite{DASH-IF, CR-Low-Latency-Live} was required. CTE is one of the main features of HTTP/1.1 (RFC 7230) [20] that allows the delivery of a segment in small pieces called chunks. The fundamental reason behind this development lies in that, for legacy DASH, the origin server has to wait for an entire segment to be encoded and packaged before advertising that segment to the client. This preparation process requires at least one segment delay. With segment sizes in 2-10 s and DASH clients requiring multiple segments downloaded even before playback starts, the added delay to the latency makes low latency infeasible. In contrast, a chunk can be as small as a single frame. Therefore, it can be delivered to the client in near real-time, even before the segment is fully encoded and available on the hosting server side. 

\begin{figure}
\centering
\includegraphics[width=0.90\linewidth, trim={0 0cm 0 0},clip]{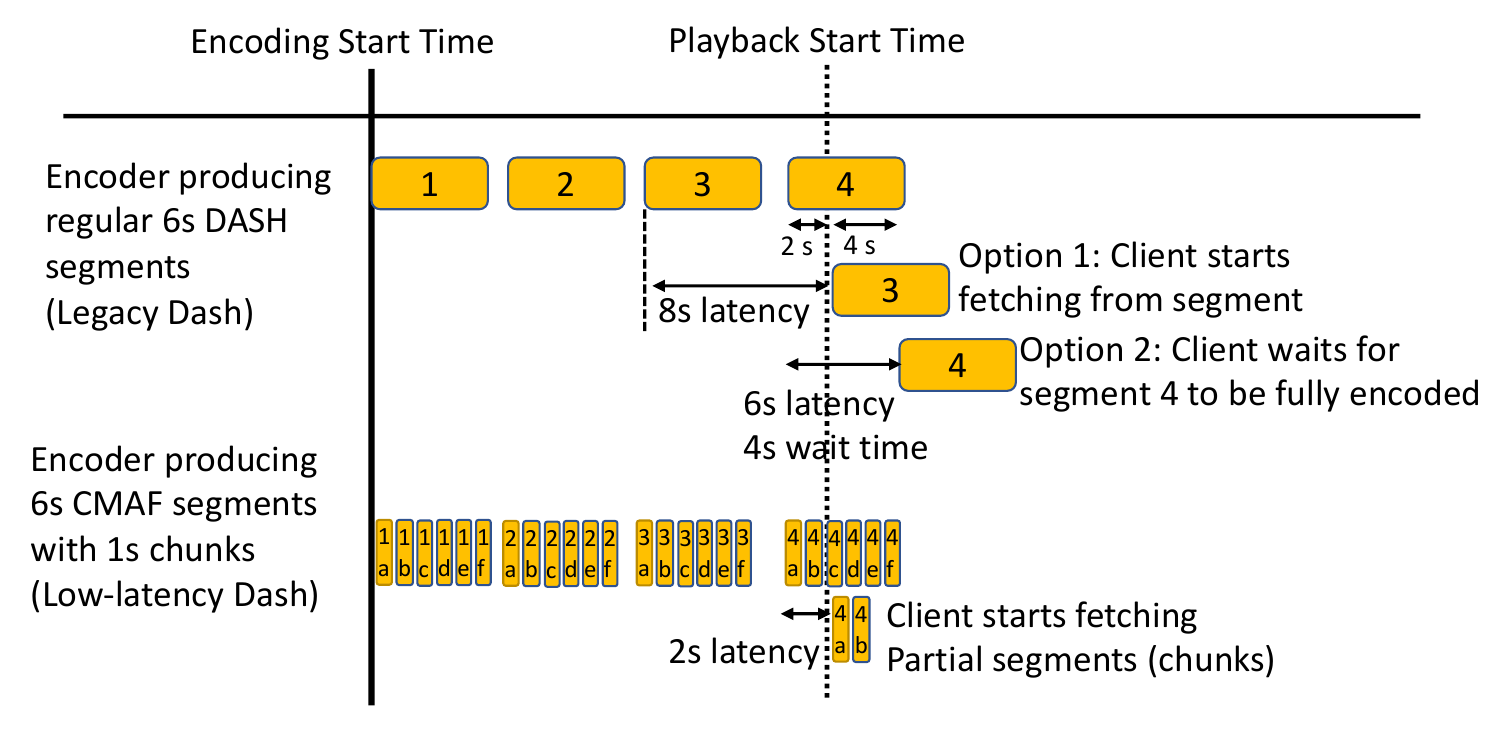}
\caption{Latency achieved with CMAF chunks.}
\label{fig:layout}
\vspace{-0.2in}
\end{figure}

Fig. \ref{fig:layout} shows the difference between Legacy DASH and Low-latency DASH when the hosting server is generating segments of 6 s length. At the current time, when playback starts, the encoder has advertised up to segment 3 and segment 4 has not been fully encoded yet. With the Legacy DASH solution, the client has two options. The first one is to skip waiting for Segment 4 and start fetching from Segment 3. In this case, the client could achieve 8 s of latency. The other option is to wait for segment 4 to be encoded fully. This way the client will achieve 6 s of latency but has to wait for additional 4 s before the video starts playing. On the contrary, for Low-latency DASH, the player can request the chunks 4a and 4b that are already advertised by the media server and starts playback as soon as receiving them, thus reducing the overall latency to 2 s.   

Additionally, to achieve low-latency streaming, per design, video clients allow a very short maximum buffer, typically in the order of 4 s. As video chunks arrive at the client, they are temporarily stored in the buffer queue to be consumed in order at a rate that equals unity, at a playback rate of 1, i.e., 1 s of video content is played back every 1 s of actual time. Therefore in a scenario where the content arrival rate is larger than 1 (requested bitrate lower than available throughput), the buffer queue would inadvertently grow, and in turn, so would the latency lag (content ages as it remains in the buffer). 

Nonetheless, shorter buffers naturally offer shorter cushions against throughput variation and/or wrong bitrate adaptation decisions. For instance, in a case of a sudden drop in the available throughput, the low-latency video client has a very short reaction window, before the buffer runs out of video data, an event that practically constitutes a stall. To recover from a stall the client is required to fill its queue up to a minimum buffer value (typically in the order of 1 s) before playback can resume -- all the while falling behind in latency. 

There are three ways to alleviate this added latency. The first pertains to requesting the most recent content from the origin server, after a stall, practically skipping a part of the video sequence with direct implications to the continuity of the streaming experience. The second is to employ a playback rate higher than 1 (speed up) until the latency lag is minimized, with implications for the streaming experience again, and with an increased probability for a "back-to-back" stall. The third way is to avoid rebuffering altogether by employing a conservative adaptation policy and a fast reaction time -- faster than the time required to register a sudden throughput drop in the throughput estimation module of typical video players (windowed approach). As we will see in Section \ref{results}, especially in scenarios with increased mobility, to avoid stalls, given the extremely short buffers employed, low-latency bit-rate adaptation requires dynamics that are refreshed faster than throughput estimates, and which can indicate a throughput drop before it even manifests. Thus, to ensure good QoE for users, it is imperative for the client device to react to dynamic network conditions. 

\vspace{-0.1in}
\section{Related Work}
 
 Several works have studied low-latency live streaming over HTTP~\cite{theo,abdelhak:nossdav2019,Stallion,Shuai:CCNC18,Hooft:journalonnsys}. 
 Bitrate adaptation is one of the key components of the low-latency streaming system and several works have tried to improve this module \cite{theo, abdelhak:nossdav2019, Stallion}. Other works aim at reducing the end-to-end latency by focusing on system configuration parameters, such as buffer size \cite{Shuai:CCNC18}, or by using very short segments over the HTTP/2 protocol \cite{Hooft:journalonnsys}.
 These works mainly use trace-driven emulation for their evaluation based on real-world traces collected mostly over WiFi or LTE networks. In contrast, our work evaluates the performance of low-latency video streaming over a real-world 5G network.
 
 There is very limited work on evaluating the end-to-end performance of low-latency video streaming over cellular networks~\cite{uplink, live5g, 7051489}. The authors in\cite{7051489} propose the use of scalable video coding for low-latency DASH over LTE networks and evaluate the performance using ns2 simulations. The works in \cite{uplink,live5g} evaluate the efficiency of low-latency streaming in a standalone 5G network testbed, focusing on uplink congestion in 5G networks. Another recent work \cite{5GAware} proposes new 5G-aware mechanisms for video streaming over 5G but focuses on volumetric video streaming over 5G networks, which has very different characteristics compared to low-latency video streaming. 
 
 
 
 
 Overall, the performance of Low-latency live streaming over real-world commercial 5G networks is mostly unknown. Through this work, we want to understand the impact of server-side configurable parameters (segment size, chunk size, and target latency selection), client-side parameters (i.e., buffer size), and lower-layer metrics (i.e., signal strength, network load) on user perceived QoE.

\vspace{-0.1in}

\section{Methodology}

\begin{figure}[!htbp] 
  \centering
    \begin{subfigure}{.4\textwidth}
        \centering
        \includegraphics[width=\textwidth]{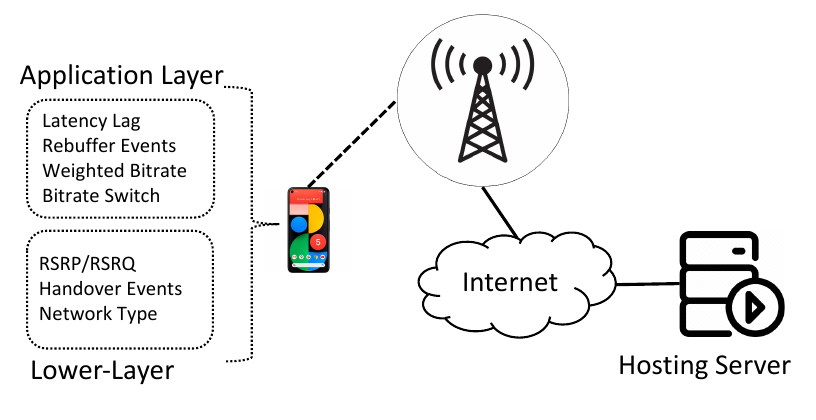}
        \label{fig:meas_setup} 
    \end{subfigure}
\vspace{-0.25in}    
\caption{Measurement setup}
\label{fig:mes_setup}
\vspace{-0.2in}
\end{figure}

In this section, we describe our experimental setup and the data collection methodology. Our setup is shown in Fig. \ref{fig:mes_setup}. The video sequence, used for live streaming, was hosted at an origin server, which performed all the transcoding and packaging operations along with serving content requests. A state-of-the-art low-latency mobile video client ran on a mobile device (smartphone) connected to the cellular network of a large US operator. 

All data collection, at both the application layer and lower-layers, took place on the mobile device. At the lower-layers, we collected cellular network throughput, radio signal metrics, and handover information, while at the application layer, we collected streaming performance metrics including video bitrate, rebuffering events, and latency. 

\subsection{Server and Content Configuration}
Our server is a Dell Precision 3340 with an Intel core i7-10700 CPU and 16GB of RAM running Ubuntu 20.04 LTS. For video transcoding and packaging, we employed a widely used and open-source video processing tool, FFMPEG \cite{FFmpeg}, that supports Low-latency DASH. 
We used the default encoder (libx264) to encode the test  video sequence at 6 different bitrates and resolutions following industry standards~\cite{encodingstats} as shown in 
Table \ref{tab:table_encoding_abel}. 
The generated segments from the FFMPEG were hosted on an HTTP/1.1 server that enables chunked transfer encoding.  

For content preparation, we focused on segment duration, chunk duration, and target latency. Different configurations of these parameters can have a significant impact on the performance of the low-latency stream. For example, the longer the segment size, the larger the target latency. Additionally, the shorter the chunk, the shorter the expected achieved latency, but that comes as a cost with the packager on the server and decoder on the client. Also, with shorter chunk duration, there can be more idle periods between the partial segments, which can increase the likelihood of inaccurate bandwidth estimation at the client. Thus, to study the inter-dependencies between the segment and chunk durations and the target latency, and 
to provide experimental insights into the tradeoffs involved, we considered a total of 9 profiles as listed in Table \ref{tab:table_label} following the recommendation from \cite{ChangingDefaultSegmentSize}. 




\begin{table}[ht]
    \centering
    \caption{Encoding Profiles.}
    \label{tab:table_encoding_abel}
    {\small
    \resizebox{1\columnwidth}{!}{\begin{tabular}{|c |c |c |c |c |c |c |c|}
    \hline
    \textbf{Resolution} & 1080P (HQ) & 1080P & 720P & 540P & 432P & 360P & 270P  \\ \hline
    \textbf{Bitrate (Mbps)}  & 6 & 4.5 & 3 & 2 & 1.1 & .73 & .365 \\ \hline
    \end{tabular}}}
    \vspace{-0.2in}
\end{table}

\begin{table}[ht]
    \centering
    \caption{Streaming Profiles. S2-C0.1-L3 denotes segment size of 2 s, chunk size of 0.1 s, and target latency  of 3 s.}
    \label{tab:table_label}
    {\small
    \begin{tabular}{|c   |c   |}
    \hline
    \textbf{Profile}             & \textbf{Segment (s)-Chunk (s)-Target Latency(s)} \\ \hline
    1,2,3   & S2-C0.1-L3, S2-C0.5-L3, S2-C1-L3 \\ \hline
    4,5,6   & S3-C0.1-L4, S3-C0.5-L4, S3-C1-L4  \\ \hline
    7,8,9   & S4-C0.1-L5, S4-C0.5-L5, S4-C1-L5 \\ \hline
    \end{tabular}}
    \vspace{-0.1in}
\end{table}

\subsection{Client Configuration and Data Collection}
We used Google Pixel 5 as the user equipment, and Google's open-source ExoPlayer \cite{ExoPlayer} version 2.16.3, which has native support for low latency, as the video client. While other video clients were also considered, such as DASH-IF's web-browser-based dash.js \cite{dashjs}, we chose Exoplayer 
since it is offered as a standalone smartphone application and typically serves as the foundational software for most of the video-based Android applications. We modified the source code of the player to change the default buffer settings (initial buffer duration for playback, default minimum buffer, default max buffer, and default buffer for playback after rebuffering) to make it suitable for the low-latency setup. We set the min and max playback rates of the player to 0.98 and 1.04 respectively. We also modified the default bandwidth estimates on the player to match them with the average throughput of the respective networks (WiFi/5G). Next, we implemented listener functions provided by the player to collect different QoE metrics. A detailed list of the metrics is given in Section \ref{results}.

To correlate how the video-client estimated bandwidth is affected by lower-layer metrics, we used G-Net track Pro \cite{G-NetTrack}, a network-based tool for collecting lower-layer information. We collected handover information, signal strength and quality (RSRP, RSRQ, RSSI), and network technology type with a granularity of 1 s.



\subsection{Measurement Scenarios}
We considered three mobility scenarios: \textit{static, walking, and driving}. For static, we first evaluated the performance of low-latency video streaming over WiFi, where the phone was connected to the server located just one hop away through a Netgear Nighthawk x10 802.11ac router. This provided us with the baseline performance, as the channel conditions were stable and there were no other clients on the WiFi channel. For each profile in Table \ref{tab:table_label}, we conducted 10 runs. 

Next, we evaluated streaming over the 5G cellular network both during \textit{peak} and \textit{non-peak} hours. To identify peak and non-peak hours, we first identified the cells that the smartphone connects to based on its location. Then, we studied the network load of those cells (by analyzing the cell utilization pattern data provided by the operator) to determine the most congested/non-congested periods of the day. We observed that the cells are most loaded during 2-8 PM and very lightly loaded around 4-7 AM. Thus, we selected these periods to represent the peak and non-peak hours respectively, and we conducted again 10 runs for each profile.

\begin{figure}[t]
  \centering
    \begin{subfigure}{.23\textwidth}
        \centering
        \includegraphics[width=\textwidth]{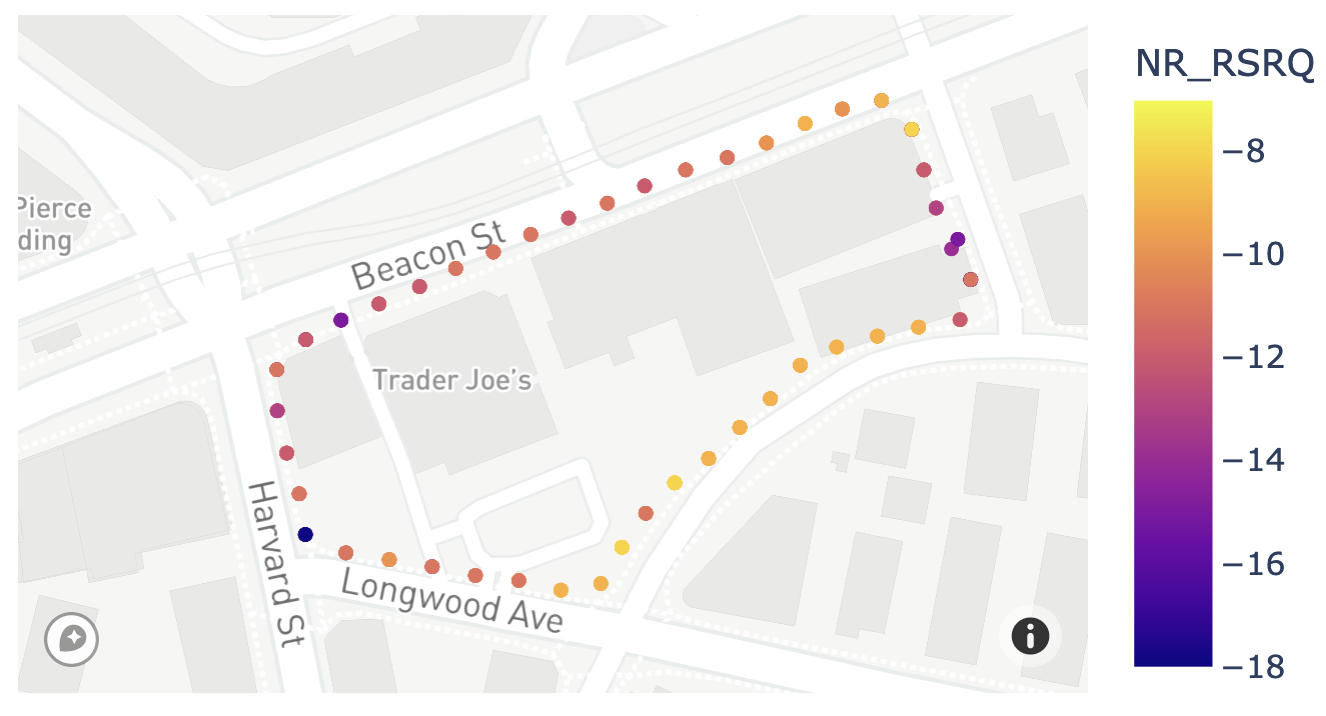}
        \caption{Walking}
        \label{fig:mobility_walk}
	\end{subfigure}
    \begin{subfigure}{.23\textwidth}
        \centering
        \includegraphics[width=\textwidth]{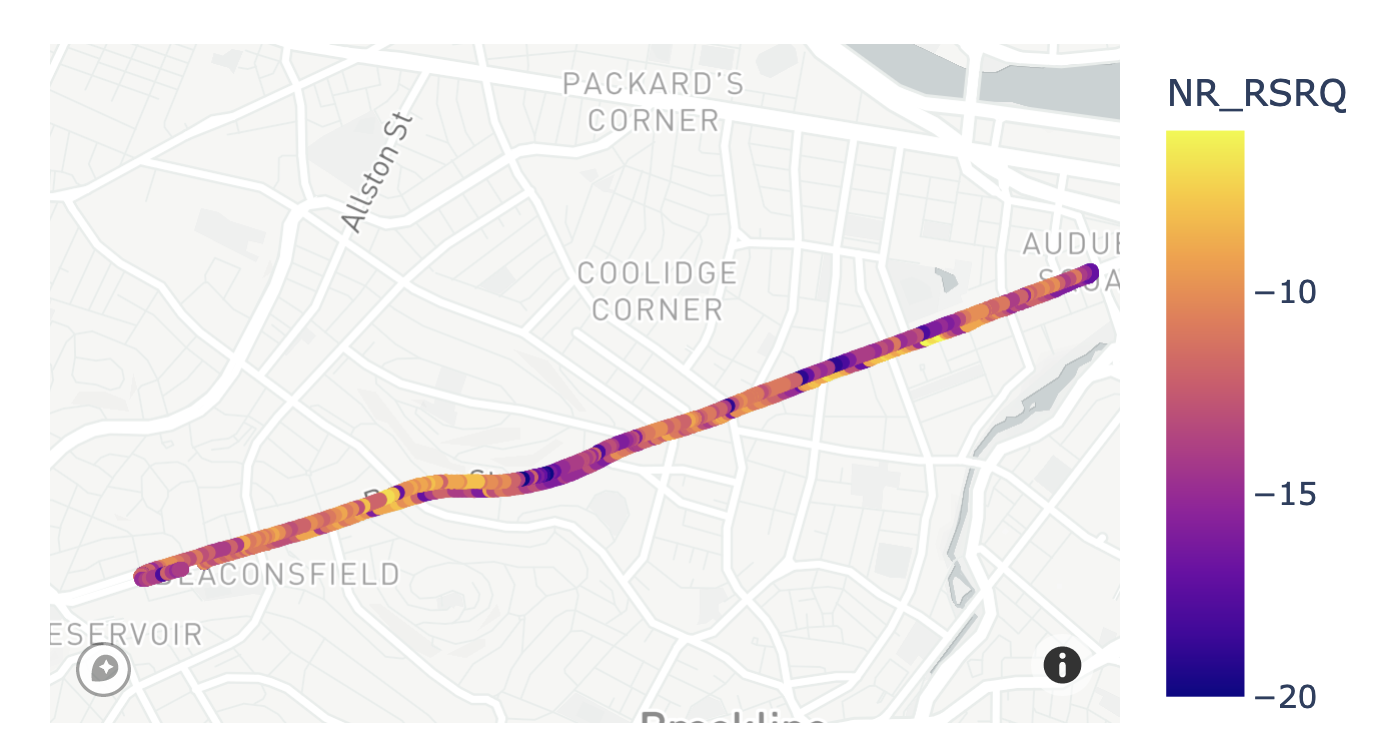}
        \caption{Driving}
	\label{fig:mobility_drive} 
    \end{subfigure}
\caption{RSRQ~[$\rm{dB}$] variation in mobility scenarios.}
\label{fig:RSRQ_Variaton}
\vspace{-0.3in}
\end{figure}

For mobility scenarios, we considered both walking and driving conditions. For walking, we walked in a loop around a base station with three cells. This ensured that a sufficient number of handovers (at least 3) and signal strength variations were observed (presented in Fig. \ref{fig:mobility_walk}). We also made sure the phone stayed on 5G most of the time (on average 90\% of the time) during playback. For driving, we similarly chose a circular route (Fig. \ref{fig:mobility_drive}), where the phone experienced a number of handover events and we ran the experiment for 20 mins for a subset of the profiles. We observed that during the driving experiments the phone experienced on average 10-15 handover events and stayed on 5G on average 95\% of the time.

\vspace{-0.1in}

\section{Measurement Results and Analysis}
\label{results}

In this section, we delve into the analysis of our low-latency streaming measurements under the three mobility scenarios.

\subsection{QoE Metrics Calculation}
We consider the following metrics for each playback session.

\noindent\textbf{Weighted Bitrate.} 
For each video playback session, the player switches between bitrates to adapt to varying network conditions. Depending on how long the player stayed on a particular bitrate, we calculate the weighted bitrate as follows:

\begin{equation}
    \rm{Weighted~Bitrate} = \frac{\sum ( b_i * \Delta t_i)}{\sum \Delta t_i}
\end{equation}
where $b_i$ is a video bitrate and $\Delta t_i$ is the duration of time during which the video bitrate was $b_i$.

\noindent\textbf{Latency Lag.} We calculate the achieved latency by averaging the per-second latency data extracted from the ExoPlayer logs. We then subtract the average value from the target value set at the server. This metric denotes the difference between achieved  and target latency.

\noindent\textbf{Bitrate Switch Count.} We count the total number of switches in bitrate observed during the playback.

\noindent\textbf{Rebuffer Event Count.} We count the number of rebuffering events recorded.

\begin{figure}[h]
  \centering
  \includegraphics[width=0.35\textwidth]{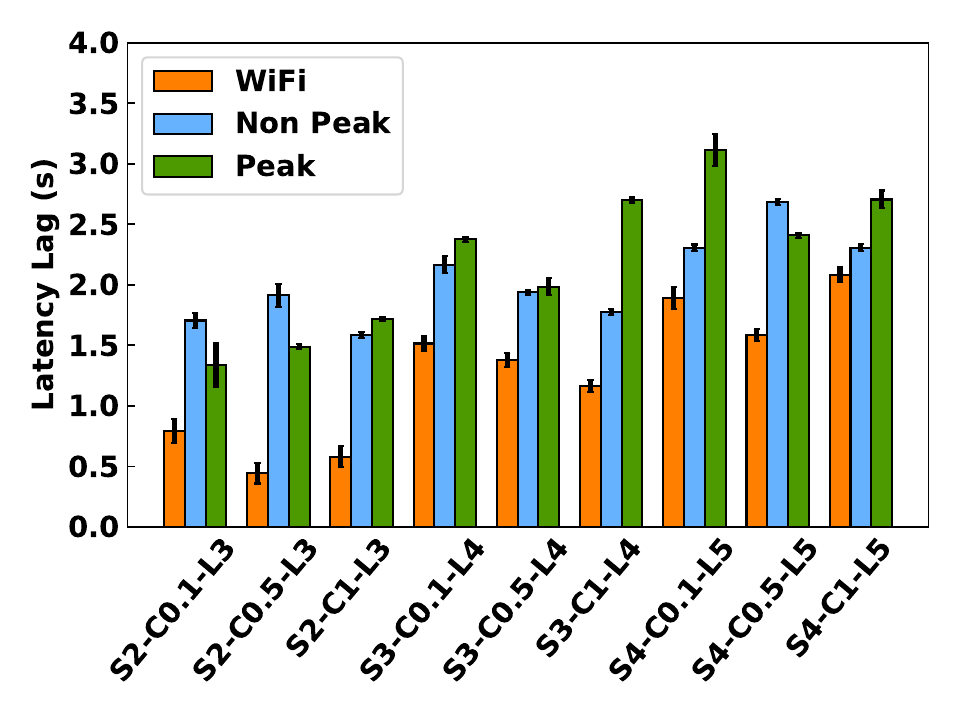}
  \vspace{-0.1in}
  \caption{Latency Lag in Static Scenario.}
  \label{fig:wifi_rates}
   \vspace{-0.3in}
\end{figure}


\begin{table*}[ht]
\centering
    \caption{Weighted Bitrate (W.B.) (Mbps) \& Rebuffer Event count (R.E.C.) under static scenario.}
    \label{tab:table_wb}
  \begin{tabular}{lSSSSSS}
    \toprule
    \multirow{2}{*}{\textbf{Profile}} &
      \multicolumn{2}{ c}{\textbf{WiFi}} &
      \multicolumn{2}{ c}{\textbf{5G (Non-peak)}} &
      \multicolumn{2}{ c}{\textbf{5G (Peak)}} \\
      & {W.B.} & {R.E.C.} & {W.B.} & {R.E.C.} & {W.B.} & {R.E.C.} \\
     \midrule
    S2-C0.1-L3 & 6.00 $\pm$ 0 & 0 & 5.87 $\pm$ 0.37 & 0 & 5.75 $\pm$ 0.47 & 0.4 $\pm$ 0.68 \\
    S2-C0.5-L3 & 6.00 $\pm$ 0 & 0 & 5.88 $\pm$ 0.24 & 0.3 $\pm$ 0.48 & 5.35 $\pm$ 1.12 & 0 \\
    S2-C01-L3 & 6.00 $\pm$ 0 & 0 & 5.69 $\pm$ 0.55 & 0.1 $\pm$ 0.32 & 5.75 $\pm$ 0.75 & 0 \\
     \midrule
    S3-C0.1-L4 & 6.00 $\pm$ 0 & 0 & 5.81 $\pm$ 0.96 & 0.1 $\pm$ 0.32 & 6.00 $\pm$ 0 & 0.2 $\pm$ 0.42 \\
    S3-C0.5-L4 & 6.00 $\pm$ 0 & 0 & 5.9 $\pm$ 0.38 & 0 & 5.84 $\pm$ 0.47 & 0.2 $\pm$ 0.42 \\
    S3-C01-L4 & 6.00 $\pm$ 0 & 0 & 6.00 $\pm$ 0 & 0.1 $\pm$ 0.32 & 5.82 $\pm$ 0.53 & 0.1 $\pm$ 0.32 \\
     \midrule
    S4-C0.1-L5 & 6.00 $\pm$ 0 & 0 & 6.00 $\pm$ 0 & 0 & 5.55 $\pm$ 0.64 & 0.4 $\pm$ 0.7 \\
    S4-C0.5-L5 & 6.00 $\pm$ 0 & 0 & 6.00 $\pm$ 0 & 0 & 5.76 $\pm$ 0.66 & 0.1 $\pm$ 0.32 \\
    S4-C01-L5 & 6.00 $\pm$ 0 & 0 & 6.00 $\pm$ 0 & 0 & 5.86 $\pm$ 0.33 & 0.2 $\pm$ 0.42 \\
   \bottomrule
  \end{tabular}
  \vspace{-0.2in}
\end{table*}

\subsection{Static Scenario}
We first consider the static scenario with WiFi and 5G (Peak and Non-Peak). The results are shown in Table \ref{tab:table_wb} and Fig. \ref{fig:wifi_rates}. We treat the Wifi scenario as our baseline, given that it presents ideal streaming performance for all content profiles, as demonstrated in the high and stable weighted bitrate and no rebuffering in Table \ref{tab:table_wb}. It is worth noting that a rebuffering event can have a large impact on latency, as our discussion will indicate in the following. While indeed our content preparation workflow is configured with tight latency targets (i.e., 3, 4, and 5 s for the different profiles evaluated), and in spite of employing a state-of-the-art video client in our experimental setup, latency may still deviate from the target, considering especially the dynamics between the origin server and the low-latency client, as explained in Section \ref{sec:background}.

In particular, we observe in Fig. \ref{fig:wifi_rates} that, even with minimal mobility and good networking conditions (static WiFi scenario, no stalls, see Table \ref{tab:table_wb}), the latency lag varies between 0.5 and 2.0 s, depending on the different content profiles. This is attributed primarily to non-network-related inter-dependencies, such as content-encoding and packaging complexities at the origin server.  Additionally,  we observe that for WiFi, the latency lag increases as a function of the segment duration (2, 3, 4 s)  and that, at least for S2 and S4, a chunk size of 0.5 s seems to be the sweet spot - an insight that could be used to inform design decisions of video-service developers.  

For 5G, we observe a higher average latency lag on Non-peak than on WiFi, while Peak shows the highest lag in almost all cases. In parallel, and according to Table \ref{tab:table_wb}, the client appears to perform adaptation, especially in the profiles with shorter segments for the case of non-peak, while all profiles show some adaptation in the peak case, irrespective of the segment duration. The more frequently observed instances of adaptation, along with the higher latency lag indicate that the Peak hours represent somewhat more challenging network conditions for low-latency streaming. In non-Peak hours, there is less cross-traffic to compete with our streaming session in the RAN, and the client manages to achieve a more stable  average bitrate relative to the case of Peak overall, while at the same time, rebuffering is less frequent at non-peak vs. peak across profiles. Nonetheless, both peak and non-peak, despite adaptation, manage to maintain a high average bitrate overall. 

Another observation concerns the relationship between rebuffering events and latency lag. Specifically, shorter segments allow more reactivity in adaptation (which, as a reminder occurs at the segment level, as opposed to the chunk level) and thus provide more flexibility toward avoiding stalls and thus preserving the latency target. Therefore, especially in the case of peak hours, we observe that the shorter segment choices serve favorably low latency. Nonetheless, in addition to using smaller segments, modern-day low-latency adaptation algorithms would require additional dynamics as inputs, to become more reactive, as will become even clearer in the discussion of our mobility results in Section \ref{sec:mobility}.

Overall, in the case of the static scenario, we are drawing two main conclusions. First, the performance of low-latency streaming depends heavily both on the network conditions and on the configuration of the video player and the origin server. It is imperative to use the right combination of segment and chunk duration to configure the content at the origin server, as well as set a viable latency target that would perhaps need to be tailored to the expected network conditions (or based on network load) where the streaming will occur. This should be one of the primary considerations for video content providers and video engineers/developers, and perhaps an interesting area for collaboration between network operators and content providers, that could focus on setting up a communication framework to exchange load information to guide latency target decisions (dynamic low-latency). Second, in our study, we do not observe large performance differences in terms of average bitrate and rebuffering when comparing peak and non-peak hours, which implies that low-latency streaming is able to support much higher encoding bitrates than the highest encoded bitrate of 6 Mbps studied in this work -- typically sufficient for encoding a 1080p resolution (AVC codec), as long as the video clients are equipped with fast low-latency-specific adaptation modules. This signifies that today's 5G production networks are able to support 2K or 4K low-latency streaming, even at peak network load. While the small screen of modern smartphone devices will not yield tremendous perceived QoE benefit at 4K, our observation raises interest in identifying mobile low-latency use cases that would.

 \begin{figure}
  \centering
   \begin{subfigure}{0.50\textwidth}
        \centering
        \includegraphics[width=\textwidth]{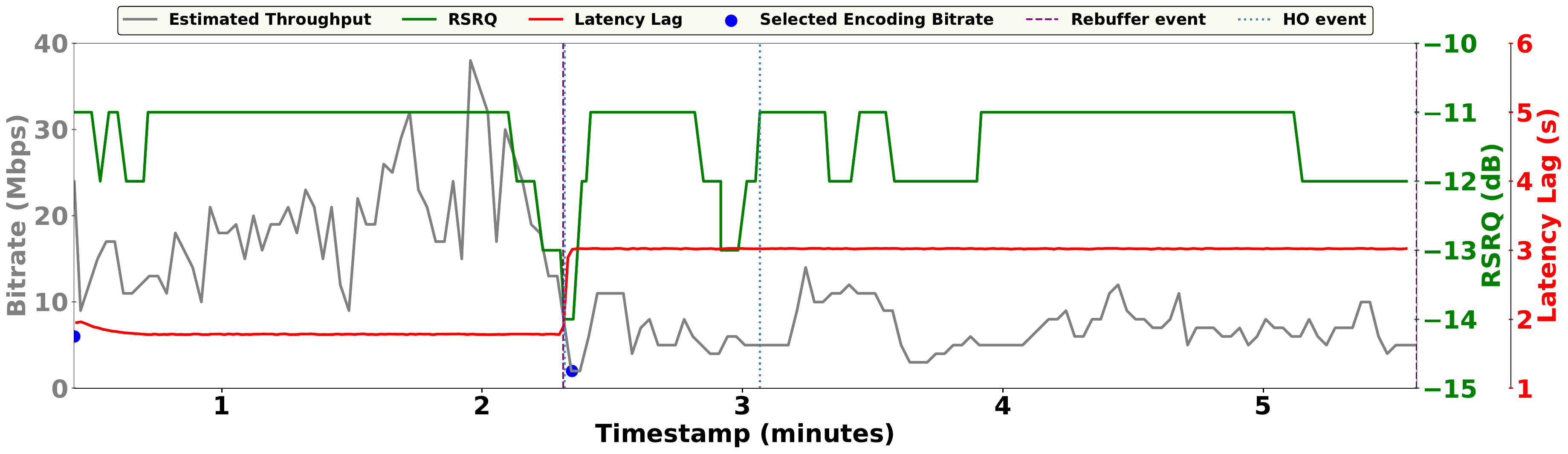}
        \caption{Sample Walking Run.}
        \label{fig:sample_walking}
	\end{subfigure}
    \begin{subfigure}{0.5\textwidth}
        \centering
        \includegraphics[width=\textwidth]{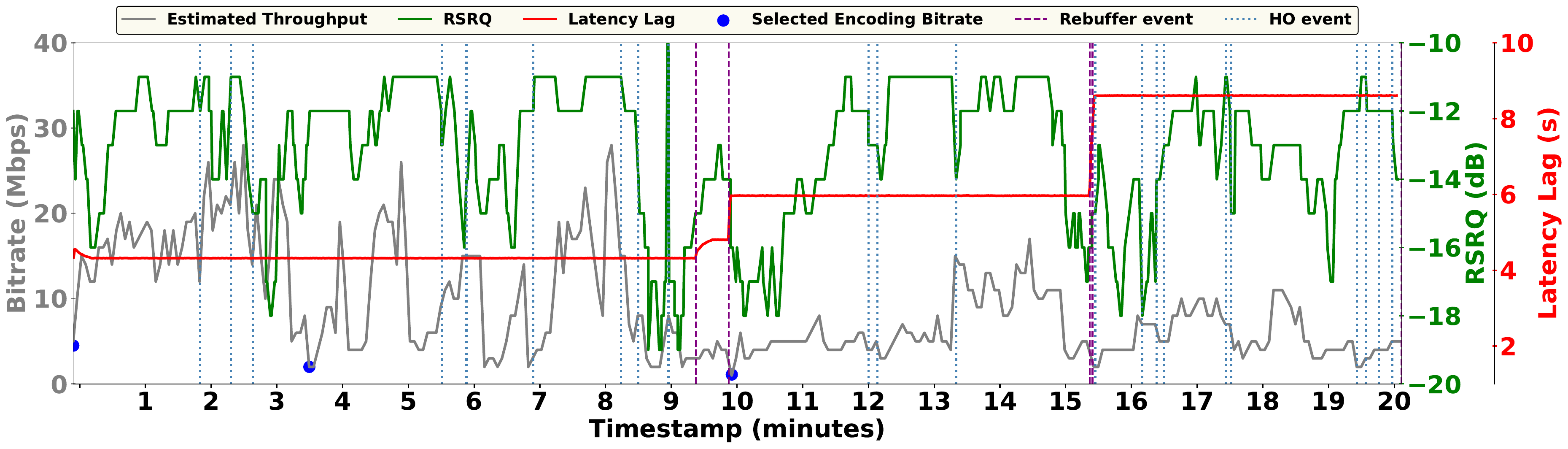}
        \caption{Sample Driving Run.}
	\label{fig:sample_driving} 
    \end{subfigure}
\caption{Mobility Scenario Case Study.}
\label{fig:sample_driving_run}
\vspace{-0.3in}
\end{figure}

        
        


\begin{figure*}[t]
  \centering
    \begin{subfigure}{.3\textwidth}
        \centering
        \includegraphics[width=\textwidth]{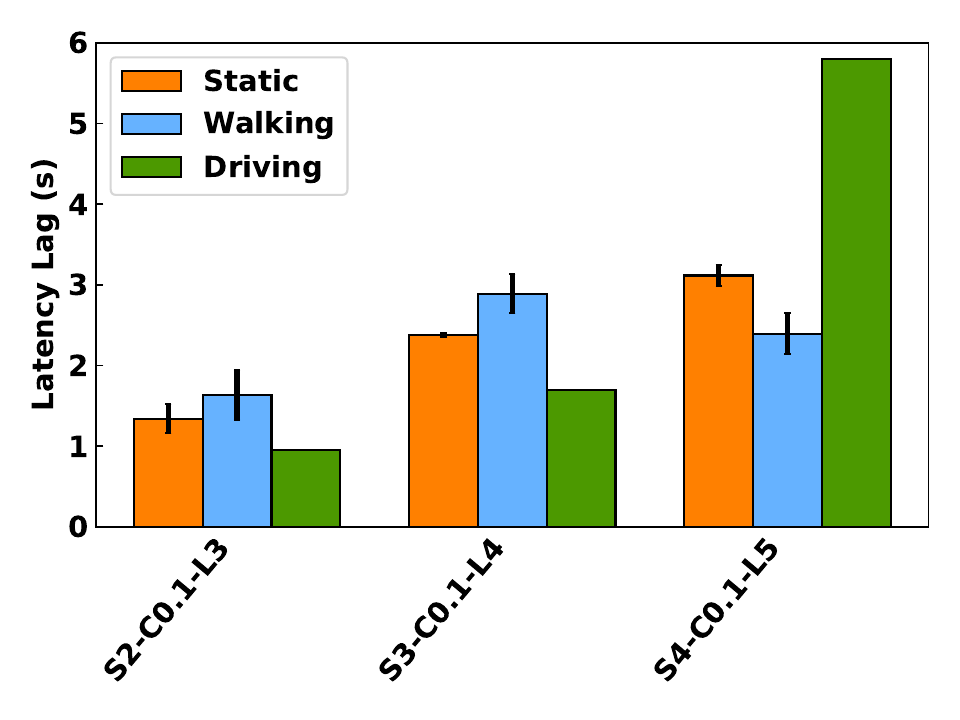}
        \vspace{-0.2in}
        \caption{Latency Lag.}
        \label{fig:comp_ld}
	\end{subfigure}
    \begin{subfigure}{.3\textwidth}
        \centering
        \includegraphics[width=\textwidth]{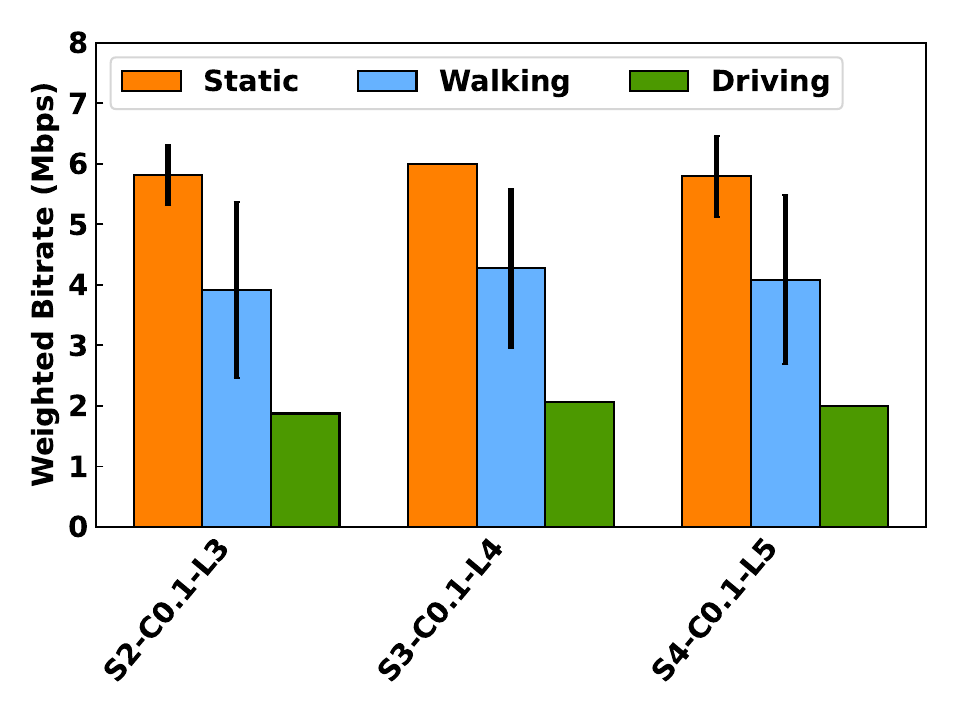}
        \vspace{-0.2in}
        \caption{Weighted Bitrate.}
	\label{fig:comp_wb} 
    \end{subfigure}
    \begin{subfigure}{.3\textwidth}
        \centering
        \includegraphics[width=\textwidth]{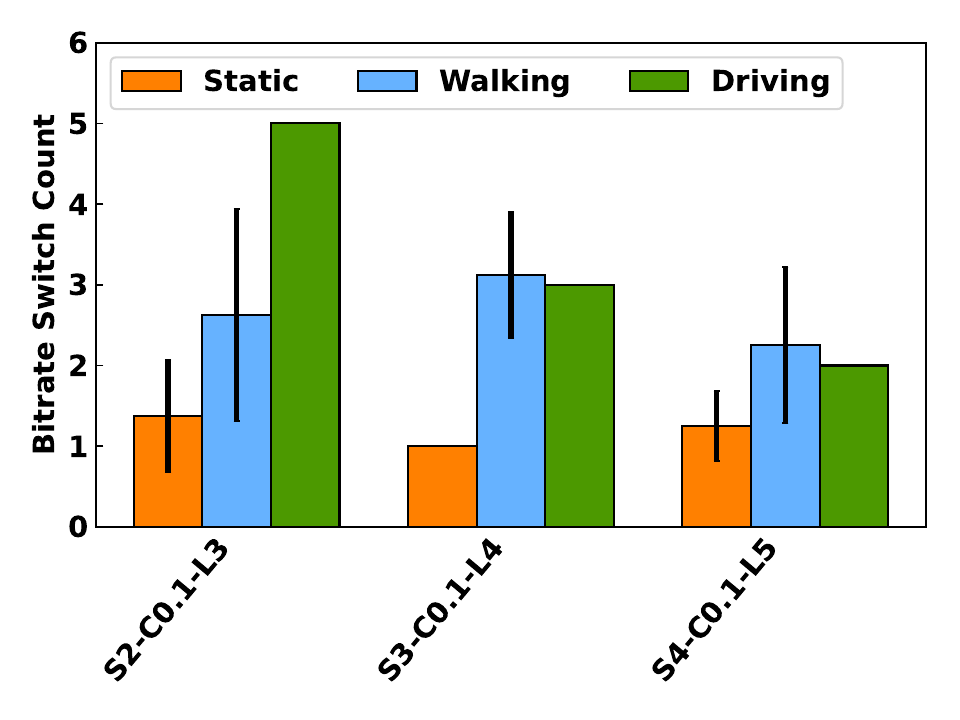}
        \vspace{-0.2in}
        \caption{Bitrate Switch Count.}
        \label{fig:comp_bsw}
    \end{subfigure}
\caption{Average QoE metrics over 5G (peak) under all mobility scenarios.}
\label{fig:scenario_compare}
\vspace{-0.2in}
\end{figure*}

\subsection{Mobility Scenario}
\label{sec:mobility}
In this section, we focus on the impact of mobility on low-latency streaming. First, we look at the timeline of one sample walking run and one driving run for the S4-C0.1-L5 profile in Fig.~\ref{fig:sample_driving_run}.  We plot the selected track bitrate, latency lag, rebuffering events, RSRQ, and handover events over time. We observe that the cell device experiences a higher number of handover events (25 vs. 2) and rebuffering events (4 vs. 1) while driving against walking. We select the particular profile  under the premise that it registers more events (handovers, rebuffering), to facilitate our discussion.  The signal strength (RSRQ) also varies significantly during the driving run. 

More specifically, for Fig. \ref{fig:sample_walking}, we observe that RSRQ presents a drop between the 2nd and 3rd minute of the playback that coincides with a handover event. While the client indeed switches to a lower encoding bitrate, it appears to not do so in time to avoid a stall. The stall event eventually causes a sharp increase in latency lag. This occurs because the client adapts based on the estimated throughput (gray line), which requires some time to register drops. Additionally, the throughput estimation is slow to respond to the network condition change. If the adaptation module was more sensitive to throughput changes, while also including RSRQ as in input signal, perhaps the stall might have been avoided. Fig. \ref{fig:sample_driving} represents a similar case, yet now due to the increased mobility while driving, handover events are much more frequent, and in turn so are stall events. In particular, between the 9th and 10th minute, we observe again the same pattern as before. A handover triggers a stall with a slight increase in latency (about 0.5-1 s) and then a back-to-back stall triggers an adaptation event, which in turn increases the lag by about 2 s. The adaptation should have occurred before either of the two stalls, especially following the abrupt drop in RSRQ.

Additionally, Fig. \ref{fig:scenario_compare} plots the QoE metrics over 5G peak hours under static, walking, and driving scenarios for three profiles. As expected, we see that, as mobility increases, so does the frequency of adaptation, which in turn results in lower average bitrate. In all cases, except for the case of driving for profile S4-C0.1-L5, the ability to down switch frequently in order to avoid rebufferings indicates a favorable impact on latency. Of course, the tradeoff comes at a lower average bitrate, yet this eventually becomes a matter of QoE prioritization, which app developers need to account for according to their specific use case (latency over bitrate, etc.).  

Overall, as indicated prior, rebuffering events are the main cause of sharp increases in latency lag. In the face of fluctuating network conditions, the short buffer associated with low latency runs out quickly causing a stall. The longer the stall the further the achieved latency will be from the target value. That effect, in conjunction with the highly dynamic network conditions especially under mobility, constitutes timely adaptation based on appropriate signals, beyond throughput estimates, mandatory for low-latency streaming

Therefore, based on the presented results of the mobility scenarios, we conclude that RSRQ can serve as a key indicator for QoE optimization. We noticed that each rebuffer event occurs when there is a sharp drop in RSRQ (e.g., timestamp 2:20th min in Fig. \ref{fig:sample_walking}, 9th min in Fig. \ref{fig:sample_driving}). A sharp drop in RSRQ could indicate to the player when to anticipate network events such as handovers, that could negatively impact user QoE. Similarly, a stable RSRQ over time could notify the player of stable network conditions to help switch up bitrates.  

In recent years some research works have proposed intelligent ABR algorithms \cite{theo, Stallion, abdelhak:nossdav2019}. These works focus mostly on client-side metrics (throughput estimation, buffer size, bitrate switch history). However, there are no works that incorporate lower-layer metrics, such as RSRQ, into the algorithm design.

\vspace{-0.09in}

\section{Conclusion}


In this paper, we carried out a first-of-its-kind measurement study on low-latency live streaming over a commercial 5G network. We examined low-latency streaming across three scenarios (static, walking, and driving) and analyzed the impact of both server-side and client-side configurations on user quality of experience (QoE). Our extensive experiments and analysis demonstrated that 5G networks can support low-latency video streaming, while also revealing the sensitivity of QoE to network dynamics such as handover events, throughput fluctuations, and signal strength variations, particularly in mobility scenarios. We also discussed the potential use of lower-layer metrics such as RSRQ to improve the accuracy of throughput estimation and thereby enhance QoE for low-latency video streaming users.
\vspace{-0.1in}

\bibliographystyle{IEEEtran}
\bibliography{mobile}

\end{document}